# Quality of Spatial Entanglement Propagation

Matthew Reichert,[*] Xiaohang Sun, and Jason W. Fleischer[†]
*Department of Electrical Engineering, Princeton University, Princeton, New Jersey 08544, USA*

We explore, both experimentally and theoretically, the propagation dynamics of spatially entangled photon pairs (biphotons). Characterization of entanglement is done via the Schmidt number, which is a universal measurement of the degree of entanglement directly related to the non-separability of the state into its subsystems. We develop expressions for the terms of the Schmidt number that depend on the amplitude and phase of the commonly used double-Gaussian approximation for the biphoton wave function, and demonstrate migration of entanglement between amplitude and phase upon propagation. We then extend this analysis to incorporate both phase curvature in the pump beam and higher spatial frequency content of more realistic non-Gaussian wave functions. Specifically, we generalize the classical beam quality parameter $M^2$ to the biphotons, allowing the description of more information-rich beams and more complex dynamics. Agreement is found with experimental measurements using direct imaging and Fourier optics.

Entanglement is a key resource in quantum information. While entanglement in discrete variables, such as spin or polarization [1-5], forms the basis of qubits, of growing interest is entanglement in continuous variables, such as transverse spatial position and momentum. The conjugate nature of these variables underlies imaging and propagation, while their infinite-dimensional Hilbert space holds much potential for quantum computation [6-9]. Typically, the photon source for continuous-variable entanglement is spontaneous parametric down-conversion (SPDC) [10-14] but, remarkably, there have been few investigations into its amount and distribution upon propagation [15,16].

A universal metric to quantify the degree of entanglement is the Schmidt number, which is directly related to the non-separability of the state's (two) subsystems [17-19]. While interferometric measurements of the Schmidt number have been proposed [15] and demonstrated [16], such methods do not examine the manifestation of the entanglement, i.e., non-separability of amplitude or phase. Furthermore, theoretical analysis has thus far focused primarily on Gaussian spatial profiles, which are not generated experimentally.

Here, we present an analysis of the Schmidt number of realistic non-Gaussian entangled photon wave functions, explicitly revealing the migration of entanglement with propagation from amplitude to phase and back again [15]. First, we present a Schmidt decomposition of the commonly used double-Gaussian approximation for the biphoton wave function. We clearly identify amplitude and phase components and demonstrate migration between them during propagation. This migration depends on the focusing geometry of the pump used to generate the photon pairs, as its phase profile directly determines the far-field properties of the biphoton wave function. We then examine more realistic biphoton wave functions that have different propagation behavior from the ideal double-Gaussian. In particular, the higher spatial frequency content of non-Gaussian beams causes increased diffraction of the biphoton. To characterize this, we introduce an effective quantum $M^2$ factor, generalizing the quality parameter commonly used to describe classical laser beams [20]. We demonstrate good predictive capability of our model, and make comparisons to experimental measurements conducted with a single-photon-sensitive electron-multiplying CCD (EMCCD) camera [21,22].

The quantum state of an entangled-photon pair (biphoton) may be described by a wave function which propagates according to Maxwell's equations [23-26]. Assuming degenerate, collinear down-conversion and a collimated strong (classical) pump beam, the momentum space wave function is [12]

$$\Phi(\mathbf{k}_1,\mathbf{k}_2) = N'\,\text{sinc}\left(\frac{L}{4k_p}|\mathbf{q}_1-\mathbf{q}_2|^2\right)\widetilde{E}_p(\mathbf{q}_1+\mathbf{q}_2) \quad (1)$$

where $N'$ is a normalization constant, $\text{sinc}(x) = \sin(x)/x$, $L$ is the crystal thickness, $k_p$ is the wave number of the pump, $\mathbf{q}_i = k_{x,i}\hat{\mathbf{k}}_{x,i} + k_{y,i}\hat{\mathbf{k}}_{y,i}$ are the transverse components of the wave vector, and $\widetilde{E}_p$ is the spatial frequency spectrum of the pump field. The real-space biphoton wave function at the output surface of the crystal is given by the 4D inverse Fourier transform of Eq. (1), which is

$$\psi(\mathbf{\rho}_1,\mathbf{\rho}_2) = N\,\text{Ssi}\left(\frac{k_p}{4L}|\mathbf{\rho}_1-\mathbf{\rho}_2|^2\right)E_p\left(\frac{\mathbf{\rho}_1+\mathbf{\rho}_2}{2}\right) \quad (2)$$

where $N$ is another normalization constant, Ssi is the shifted sine integral [27], $\mathbf{\rho}_i = x_i\hat{\mathbf{x}}_i + y_i\hat{\mathbf{y}}_i$ are the transverse coordinates of each photon, and $E_p$ is the spatial distribution of the pump field. In general, Eq. (2) is not separable in the coordinates of the two subsystems ($\mathbf{\rho}_1$, $\mathbf{\rho}_2$), meaning it represents a spatially entangled state. However, it is separable in the sum and difference coordinates, defined by $\mathbf{\rho}_\pm = (\mathbf{\rho}_1 \pm \mathbf{\rho}_2)/\sqrt{2}$, i.e., $\psi(\mathbf{\rho}_+,\mathbf{\rho}_-) = \psi_-(\mathbf{\rho}_-)\psi_+(\mathbf{\rho}_+)$. Phyically, $\psi_+(\mathbf{\rho}_+)$ is proportional to the spatial profile of pump field and

$\psi_-(\boldsymbol{\rho}_-)$ depends on the longitudinal profile of the nonlinear susceptibility of the crystal, which we have here assumed to be constant.

Because Eq. (2) is rather inconvenient to work with, the biphoton wave function is often approximated by a double-Gaussian function [12,15,28-30], which in $(\boldsymbol{\rho}_+, \boldsymbol{\rho}_-)$ coordinates is

$$\psi_{dG}(\boldsymbol{\rho}_+,\boldsymbol{\rho}_-) = N\exp\left\{-\left(\frac{1}{4\sigma_-^2}+i\frac{k_0}{2R_-}\right)\rho_-^2 - \left(\frac{1}{4\sigma_+^2}+i\frac{k_0}{2R_+}\right)\rho_+^2 + i\zeta\right\} \quad (3)$$

where $k_0 = 2\pi/\lambda$, is the down-converted wavelength. $\psi_{dG}$ has many similarities with classical Gaussian laser beams. In particular, the standard deviations of $|\psi_{dG}|^2$ in sum and difference coordinates $\sigma_\pm$ have the same $z$ dependence as the radius of a TEM$_{00}$ Gaussian beam

$$\sigma_\pm(z) = \sigma_{\pm,0}\sqrt{1+\left(\frac{z}{z_{0,\pm}}\right)^2} \quad (4)$$

where $\sigma_{\pm,0} = \sigma_\pm(z=0)$ are the minima. The quantities $z_{0,\pm} = 2k\sigma_{\pm,0}^2$ are analogous to the Rayleigh range of a classical Gaussian beam. The spatially dependent phase terms depend on the radii of curvature along the $\boldsymbol{\rho}_\pm$ directions, which follow

$$R_\pm(z) = z\left[1+\left(\frac{z_{0,\pm}}{z}\right)^2\right] \quad (5)$$

in analogy to the radius of curvature of the phase fronts of a Gaussian beam. Finally, $\zeta(z) = \tan^{-1}(z/z_{0,-}) + \tan^{-1}(z/z_{0,+})$ is analogous to the classical Gouy phase. Based on this analytic double-Gaussian wave function, it is straightforward to evaluate entanglement during propagation.

A general approach to characterizing the degree of entanglement of a bipartite system is via the Schmidt decomposition, which expresses a pure entangled state as $|\Psi\rangle = \sum_n \lambda_n^{1/2} |u_n\rangle|v_n\rangle$, where $\sum_n \lambda_n = 1$. As the number of terms is directly related to the non-separability of the state, the degree of entanglement can be characterized by the Schmidt number $K \equiv (\sum_n \lambda_n^2)^{-1}$. While $K$ describes the fundamental meaning of entanglement, i.e., the non-separability of the two sub-systems [18,19,28], it contains no information about the form in which the entanglement is manifest—that is, in amplitude or phase [15,16]. For the double-Gaussian wave function, the Schmidt number can be expressed directly in terms of the wave function (see Supplementary Information)

$$K_{dG} = \left[\iint \psi_{dG}^*(\boldsymbol{\rho}_+,\boldsymbol{\rho}_-)\psi_{dG}(-\boldsymbol{\rho}_-,-\boldsymbol{\rho}_+)d^2\boldsymbol{\rho}_+ d^2\boldsymbol{\rho}_-\right]^{-1} \quad (6)$$

Evaluating $K_{dG}$ with Eq. (3) yields an analytic expression

$$K_{dG} = \frac{1}{4}\left(\frac{\sigma_+}{\sigma_-}+\frac{\sigma_-}{\sigma_+}\right)^2 + k^2\sigma_+^2\sigma_-^2\left(\frac{1}{R_-}-\frac{1}{R_+}\right)^2 \quad (7)$$

In sum, $K_{dG}$ represents the total degree of entanglement, and is related to the average first-order coherence of the beam [31,32]. In parts, the two terms have a clear physical meaning. The first term depends only on the amplitude of $\psi_{dG}$, specifically the ratios of the standard deviations in $\boldsymbol{\rho}_+$ and $\boldsymbol{\rho}_-$. We will therefore refer to this as $K_{amp}$. It is essentially a measure of how many correlation areas there are within the beam, and is equal to the expression from [28]. The second term depends on the phase of the biphoton wave function, in terms of the difference of curvature in the $\boldsymbol{\rho}_+$ and $\boldsymbol{\rho}_-$ planes. We therefore refer to the second term as $K_{phase}$. This represents, to the best of our knowledge, the first explicit expression for entanglement within the spatial phase of entangled photon pairs.

Fig. 1 shows the evolution of the components of the Schmidt number upon propagation, using Eqs. (4) and (5) in Eq. (7), showing migration from amplitude to phase and back as the biphoton beam goes from the near-field to the far-field. The wavelength here and throughout is 810 nm. Note that in this case $z_{0,-} \ll z_{0,+}$, and the entanglement in amplitude and phase are equal at $z = z_{0,-}$ and $z = z_{0,+}$.

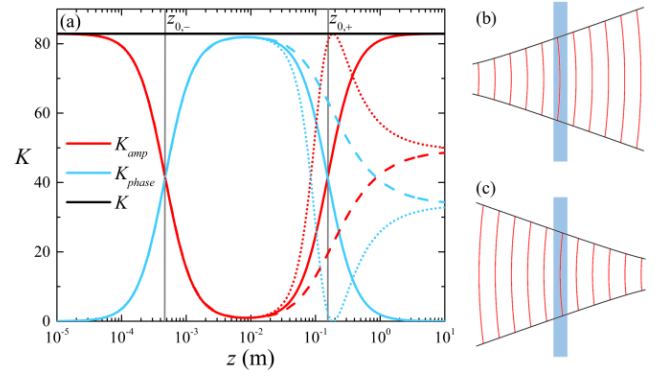

FIG. 1. Migration of spatial entanglement between amplitude and phase. (a) Evolution of the components of the $K_{dG}$ with $\sigma_+(0) = 100$ µm and $\sigma_{-,0} = 5.5$ µm for (solid) collimated, and non-collimated pump beam (dashed, b) $R_p = 18.7$ cm ($z_p = -7.6$ cm) and (dotted, c) $R_p = -18.7$ cm ($z_p = 7.6$ cm). Rayleigh ranges are indicated by the gray vertical lines indicate $z_{0,-} = 470$ µm and $z_{0,+} = 155$ mm. The Schmidt number begins in the near-field entirely in the amplitude (red (dark gray)), migrates into the phase (blue (light gray)) upon propagation, and back to amplitude in the far-field, such that the total (top black) remains constant. Rayleigh ranges $z_{0,-} = 469$ µm, and $z_{0,+} = 155$ mm are indicated by the vertical gray lines. For non-collimated pump beams, $K_{amp}$ and $K_{phase}$ follow much the same form as the collimated case up to $z \approx 2$ cm, beyond which the influence of the curved pump wavefronts causes an altered migration of entanglement, and different far field behavior

The effect of the pump beam's phase profile can also be included. A non-collimated Gaussian pump, i.e., a focusing or defocusing beam, has field profile

$$E_p(\boldsymbol{\rho}) = E_{p,0}\exp\left\{-\left(\frac{1}{4\sigma_p^2}+i\frac{k_p}{2R_p}\right)\rho^2\right\}. \quad (8)$$

where $\sigma_p$ and $R_p$ are the pump's standard deviation and radius of curvature, respectively. At the crystal, $\sigma_+ = \sqrt{2}\sigma_p$,

$R_+ = R_p$, and $z_{0,+} = z_{0,p}$ (the Rayleigh range of the pump). We define the distance between the pump beam waist and the PDC crystal as $z_p$, and incorporate its effect on $\psi_{dG}$ by letting $z \to z - z_p$ in the expressions for $\sigma_+$ and $R_+$. In Fig. 1 we also show two additional cases of the evolution of the Schmidt number: the pump focusing before and after the crystal. Here, $K_{phase}$ is nonzero in the far-field due to the phase curvature of the pump, given by the initial difference between $R_+$ and $R_-$ (at $z = 0$, $R_- = \pm\infty$ and $R_+ = R_p$). When the pump is focusing into the crystal, $R_+$ starts out negative, goes to $-\infty$ as the phase fronts flatten, changes sign to $+\infty$ at $z = z_p$ as divergence takes over, reduces to a local minimum value at $z = z_{0,+}$, and then increases. In contrast, $R_-$ starts out positive and increases faster than $R_+$, since $z_{0,-} < z_{0,+}$. At $z \approx z_p + z_{0,+}^2/z_p$ (assuming $\sigma_{-,0} \ll \sigma_{+,0}$), the curvatures become equal ($R_+ = R_-$) and $K_{phase}$ drops to zero. For positive $R_p$ (negative $z_p$), $R_+$ is always positive and greater than $R_-$, and no such oscillatory features are observed.

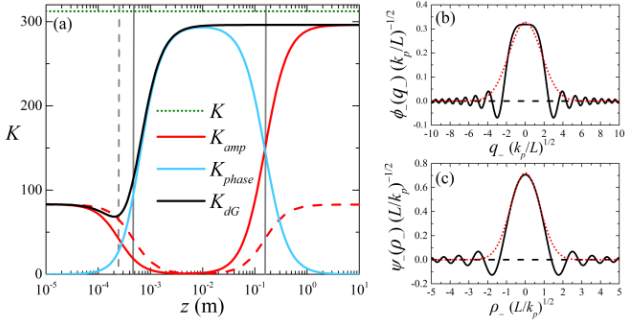

FIG. 2. Evolution of entanglement of realistic biphoton wave function. (a) Evolution of the "Gaussian" components of the Schmidt number (Eq. (7)), of the realistic biphoton wave function (Eq. (2)) where $\sigma_{-,0} = 5.5$ µm, $\sigma_{+,0} = 100$ µm, and $M^{-2} = 1.89$ which contains only a portion of (dotted green line) the total Schmidt number $K$. Rayleigh ranges are indicated by the vertical solid gray lines, and the dashed gray line is $z_{0,-} / M^{-2}$. The higher spatial frequency content of the Ssi($x^2$) function causes greater diffraction (in $\rho_-$) than the double-Gaussian (dashed red curve), leading to a more rapid migration of entanglement from amplitude to phase, and higher maxima of $K_{phase}$ and $K_{amp}$ in the far-field. Difference-coordinate dependence of the (solid black curve) realistic biphoton wave function (Eqs. (1) and (2)) with (dotted red curve) Gaussian fits in the (b) near- and (c) far-field. The oscillatory structure of the realistic wave function is lost in the Gaussian approximation. In (b) $\mathbf{q}_- = (\mathbf{q}_1 - \mathbf{q}_2)/\sqrt{2}$.

Unfortunately, the actual biphoton wave function is not well approximated by a double-Gaussian, particularly not upon propagation. FIG. 2(b) and 2(c) show the dependence of the realistic biphoton wave function on difference coordinates, along with a Gaussian fit, in both the near and far fields. The fine oscillatory structure is completely lost in the double-Gaussian approximation. $K_{amp}$ in Eq. (7) therefore does not represent the entirety of the entanglement of the amplitude. In particular, the functions may have the same variances, and thus the same $K_{amp}$, but very different Schmidt numbers $K$. The oscillatory nature of Ssi($x^2$) means that the amplitude of Eq. (2) will never be separable in ($\rho_1$, $\rho_2$) coordinates, even when $\sigma_- = \sigma_+$. This means that the common experimental practice of measuring $\{\sigma_\pm, \sigma\}$ and evaluating $K_{amp}$ [14,33] does not accurately capture the entirety of the spatial entanglement information content, but rather only a small portion [16,28].

In general, proper evaluation the Schmidt number must be done numerically, so there is no analytic form that clearly identifies the amplitude and phase components. Still, we may use Eq. (7) to evaluate the "Gaussian" components of the Schmidt number, with the understanding that they will necessarily be less than the total. To do so, we numerically propagate the realistic biphoton wave function (Eq. (2) with a collimated Gaussian pump) a distance $z$ and fit the result to a double-Gaussian wave function to determine $\sigma_\pm$ and $R_\pm$. Fitting in this way, rather than, say, directly evaluating the variances and effective radii of curvature [34], yields a Gaussian approximation that more accurately reflects both the peak probability density and its full width at half maximum (FWHM) [12]. This is essentially what is done experimentally: measure the biphoton probability distribution and fit the result to a Gaussian to determine $\sigma_-$ and $\sigma_+$ [14,16,35-37]. For $\sigma_- = 5.5$ µm and $\sigma_+ = 100$ µm, this procedure yields FIG. 2(a). This "Gaussian" part of the Schmidt number, $K_{dG}$, not only migrates from amplitude to phase and back, but also changes its total upon propagation, although never reaching the total Schmidt number $K$.

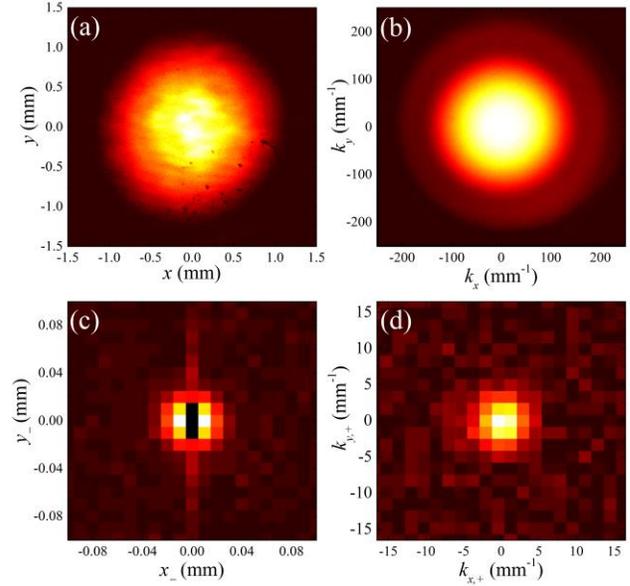

FIG. 3. Images of the (a, b) irradiance and (c, d) correlation distributions in the (a, c) near and (b, d) far field. Black regions of (c) have been zeroed out to eliminate charge-smearing artifact [14]

The evolution of the components of $K_{dG}$ of the realistic biphoton wave function can be accounted for through a modified double-Gaussian approximation. The Ssi($x^2$) function has higher spatial frequency content than a Gaussian and therefore diffracts more rapidly, resulting in a more rapid migration of entanglement from amplitude into phase. To describe such increased diffraction, we borrow the concept of "beam quality" parameter $M^2 \equiv 2\sigma_x\sigma_k$ commonly used to characterize classical laser beams [20]. It is a measure of how far a beam is from diffraction-limited, and equal to the ratio of the divergence angle of a realistic beam to that of an ideal TEM$_{00}$ Gaussian beam of the same waist. For this ideal minimum-uncertainty beam, $M^2 = 1$, while deviations (enhanced diffraction) result in increasing $M^2$. To extend this concept to the biphoton wave function, we introduce the dual "biphoton quality" parameters $M_\pm^2$, in $\rho_-$ and $\rho_+$ coordinates. Using the Gaussian fits in FIG. 2(b) and 2(c) of the realistic biphoton wave function $\psi_-$, we find $M_-^2 = 1.89$, while $M_+^2 = 1$ (since the pump is assumed to be TEM$_{00}$). We can then modify both $\sigma_\pm(z)$ and $R_\pm(z)$ in the same manner done for classical laser beams by modifying the Rayleigh ranges $z_{0,\pm} \to z_{0,\pm}/M_\pm^2$. Thus Eqs. (4) and (5) become

$$\sigma_\pm(z) = \sigma_{\pm,0}\sqrt{1+\left(M_\pm^2\frac{z}{z_{0,\pm}}\right)^2}, \quad (9)$$

and

$$R_\pm(z) = z\left[1+\left(\frac{1}{M_\pm^2}\frac{z_{0,\pm}}{z}\right)^2\right], \quad (10)$$

respectively. Evaluating $K_{amp}$ and $K_{phase}$ using this modified expression reproduces the "Gaussian" parts of the Schmidt number in FIG. 2(a).

An interesting point is that the ratio of $K_{amp}$ in the near and far fields is related to that of $M_+^2$ and $M_-^2$. Assuming, $\sigma_{-,0} \ll \sigma_{+,0}$ (or alternatively $\sqrt{(L/k_p)} \ll \sigma_p$), as is typically the case in experiment, we find

$$\frac{K_{amp}(z\to\infty)}{K_{amp}(z=0)} \approx \left(\frac{M_-^2}{M_+^2}\right)^2. \quad (11)$$

For an ideal double-Gaussian wave function, where $M_\pm^2 = 1$, the ratio is unity, while for the realistic biphoton wave function in Eq. (2) it is 3.56 (see FIG. 2(a)). This agrees well with the value calculated from Eq. (11) of $(1.89/1.0)^2 = 3.57$.

To confirm this behavior, we performed experiments using an electron multiplying CCD (EMCCD) camera, which has both single-photon sensitivity and massively parallel measurement capabilities, making it convenient for biphoton measurements [14,37-39]. A spatially filtered 405 nm CW laser beam pumps a type I SPDC crystal (BBO, $L = 3$ mm), generating near-collinear entangled photons, and nearly degenerate pairs are selected via spectral filter. Two lens systems image the near- and far-fields of the crystal onto the camera, which has 16×16 μm$^2$ pixels, and operates at −85 °C, a 17 MHz read out rate, and 0.3 μs vertical shift time. The marginal (irradiance) distributions are measured by long exposures. To measure the conditional distributions, the camera is operated in photon counting mode, with a binary thresholding on each pixel level and a mean count rate per pixel of ~0.15, chosen to optimize the signal-to-noise ratio [40]. Conditional probabilities are calculated by auto-correlation (self-convolution) of each frame for near-field (far-field), with background subtracted, calculated via cross-correlation (cross-convolution) between successive frames. $10^4$ frames were collected at each $z$-position. The camera was translated $\Delta z$ about $z = 0$, and for far-field measurements the effective far-field distance was calculated using $z = f(f + m^2\Delta z)/(m^2\Delta z)$, where $f$ is the focal length of the Fourier transform lens and $m$ is the magnification.

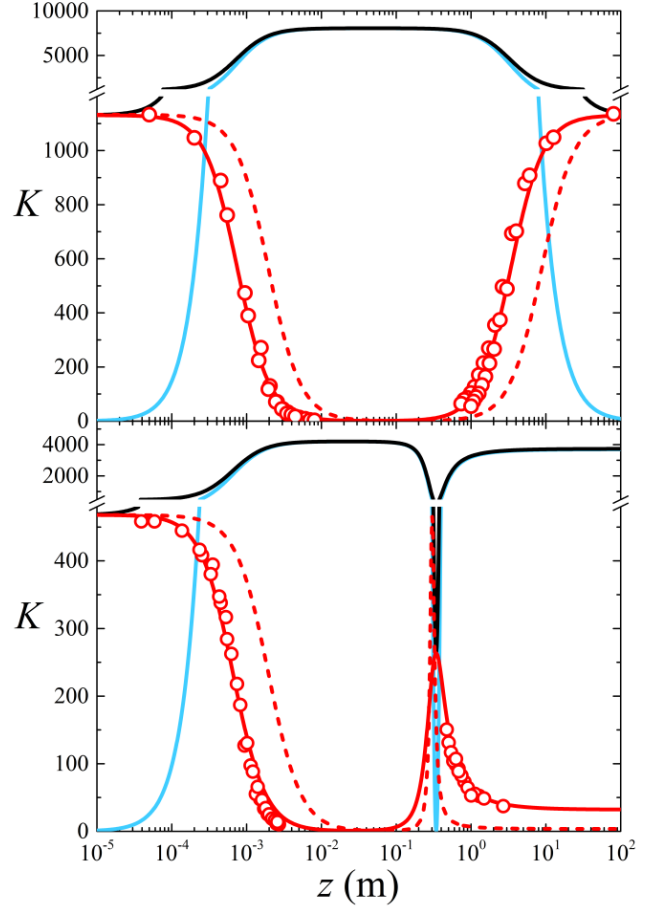

FIG. 4. Comparison of (circles) measured $K_{amp}$ with (solid red (dark gray)) theory for (a) $\sigma_- = 11.2$ μm, $\sigma = 533$ μm, $M_+^2 = M_-^2 = 2.67$, and $z_p = 0$, and (b) $\sigma_- = 11.3$ μm, $\sigma = 345$ μm, $M_+^2 = 4$, $M_-^2 = 3$, and $z_p = 30$ cm ($R_p = -96$ cm). Dashed red curve shows $K_{amp}$ with same parameters but $M_\pm^2 = 1$. Blue (light gray) and top black curves show corresponding $K_{phase}$ and $K_{dG}$, respectively. Errors in measured $K_{amp}$ are ~10 % (not shown).

FIG. 3 shows typical measurements in both the near- and far-fields. At these planes, both the spatial correlation in real-space (position) and anti-correlation in $k$-space (momentum) are approximately Gaussian. Their relative uncertainties are given by their standard deviations; correspondingly, it is easy to demonstrate the EPR paradox. In our case, we find a violation $\sigma_{x-}\sigma_{k+} = (2.7 \pm 0.1) \times 10^{-2} \ll 1/2$. In terms of information content, we find $K_{amp} = 1133 \pm 38$ in the near field and $1136 \pm 70$ in the far field. The irradiance and correlation profiles for several defocused planes from the ideal image and Fourier planes of the crystal reveal the fall-off of the correlation with propagation, i.e., the decay of entanglement in amplitude.

FIG. 4 shows measurements of $K_{amp}$ about the near and far fields, for both collimated and focusing pump beams. Excellent agreement with theory is obtained with values of $M_\pm^2 > 1$. Also plotted are the corresponding components $K_{phase}$ and $K_{dG}$, which reach values nearly an order of magnitude greater than $K_{amp}$, indicating that only a small portion of the total Schmidt number resides in the "Gaussian" part. Note that many effects could be responsible for the relatively large values of $M_\pm^2$, such as the bandwidth of the bandpass filters (50 nm), non-collinear phase matching, spatial walk-off within the crystal, aberrations in the imaging systems, and imperfect spatial filtering.

In conclusion, we have studied the dynamics of spatial entanglement and the distribution of information content with the biphoton wave function. For the double-Gaussian approximation for entangled photon pairs, we have presented an analytic expression of the Schmidt number with separate amplitude and phase terms, and studied migration of entanglement between the two with propagation. For more realistic wave functions, we introduced the biphoton quality parameters $M_\pm^2$ to allow more accurate modeling and better characterization of the evolution of spatial entanglement. By identifying the information content in the amplitude and phase, and following the migration between them, it becomes possible to engineer the degree of entanglement in either component and transfer it to the other upon propagation. With more parameters, these ideas can be extended to include higher-order moments and address more degrees of freedom within the fine structure of continuous-variable wave functions.

## ACKNOWLEDGMENTS

This work was supported by AFOSR grants FA9550-14-1-0177 and FA9550-12-1-0054.

## APPENDIX

We derive the expression for the "Gaussian" part of the Schmidt number in terms of the biphoton wave function, equation (6) in the main text. Following [16], we can express the Schmidt modes in terms of the creation operators of the signal and idler photons

$$|\Psi\rangle = \sum_n \sqrt{\lambda_n} \hat{A}_n^\dagger \hat{B}_n^\dagger |0,0\rangle \quad (A12)$$

where

$$\hat{A}_n^\dagger |0\rangle = \int \varphi_n(\boldsymbol{\rho}_1) \hat{a}^\dagger(\boldsymbol{\rho}_1)|0\rangle \, d^2 \boldsymbol{\rho}_1, \quad (A13)$$

$$\hat{B}_n^\dagger |0\rangle = \int \varphi_n(\boldsymbol{\rho}_2) \hat{a}^\dagger(\boldsymbol{\rho}_2)|0\rangle \, d^2 \boldsymbol{\rho}_2. \quad (A14)$$

With the electric field operator

$$\hat{E}^{(+)}(\boldsymbol{\rho}_1) = \sum_n \hat{A}_n^\dagger \varphi_n^*(\boldsymbol{\rho}_1) \quad (A15)$$

and $\hat{E}^{(-)}(\boldsymbol{\rho}_1) = [\hat{E}^{(+)}(\boldsymbol{\rho}_1)]^\dagger$, the first-order coherence function is given by

$$G^{(1)}(\boldsymbol{\rho}_1, \boldsymbol{\rho}_1') = \sum_n \lambda_n \varphi_n(\boldsymbol{\rho}_1) \varphi_n^*(\boldsymbol{\rho}_1'), \quad (A16)$$

and likewise for $\boldsymbol{\rho}_2$. The Schmidt modes of a double-Gaussian wave function are Hermite-Gaussian polynomials, which have symmetry properties

$$\varphi_{2n}(-\boldsymbol{\rho}_1) = \varphi_{2n}(\boldsymbol{\rho}_1), \quad (A17)$$

$$\varphi_{2n+1}(-\boldsymbol{\rho}_1) = -\varphi_{2n+1}(\boldsymbol{\rho}_1). \quad (A18)$$

Letting $\boldsymbol{\rho}_1' = -\boldsymbol{\rho}_1$ allows simplification of the first-order coherence function

$$G_{dG}^{(1)}(\boldsymbol{\rho}_1, -\boldsymbol{\rho}_1) = \sum_n \left( \lambda_{2n} |\varphi_{2n}(\boldsymbol{\rho}_1)|^2 - \lambda_{2n+1} |\varphi_{2n+1}(\boldsymbol{\rho}_1)|^2 \right)$$
(A19)

where we have added the subscript $dG$ to indicate that this is valid for the "double-Gaussian" wave function. Integrating over $\boldsymbol{\rho}_1$, and using the normalization of $\varphi_n(\boldsymbol{\rho}_1)$, yields

$$\int G_{dG}^{(1)}(\boldsymbol{\rho}_1, -\boldsymbol{\rho}_1) d^2 \boldsymbol{\rho}_1 = \sum_n (\lambda_{2n} - \lambda_{2n+1}). \quad (A20)$$

We now apply two properties of the eigenvalues of the Schmidt modes. First; they decrease exponentially with $n$, i.e., $\lambda_n = \lambda_0 \alpha^{-n}$ [15,16,30,41], yielding

$$\int G_{dG}^{(1)}(\boldsymbol{\rho}_1, -\boldsymbol{\rho}_1) d^2 \boldsymbol{\rho}_1 = \sum_n \lambda_0 \left( \alpha^{-2n} - \alpha^{-(2n+1)} \right)$$
$$= \sum_n \lambda_0 (1 - \alpha^{-1}) \alpha^{-2n}. \quad (A21)$$

Second; they are normalized such that $\sum_n \lambda_n = 1$, meaning $\lambda_0 = 1 - \alpha^{-1}$. This allows the simplification

$$\int G_{dG}^{(1)}(\boldsymbol{\rho}_1, -\boldsymbol{\rho}_1) d^2 \boldsymbol{\rho}_1 = \sum_n (\lambda_0 \alpha^{-n})^2 = \sum_n \lambda_n^2, \quad (A22)$$

the inverse of which is the definition of the Schmidt number. Therefore the "Gaussian" part of the Schmidt number is related to the average first-order coherence.

In terms of the double-Gaussian biphoton wave function

$$G_{dG}^{(1)}(\boldsymbol{\rho}_1, \boldsymbol{\rho}_1') = \int \psi_{dG}^*(\boldsymbol{\rho}_1, \boldsymbol{\rho}_2) \psi_{dG}(\boldsymbol{\rho}_1', \boldsymbol{\rho}_2) d^2 \boldsymbol{\rho}_2, \quad (A23)$$

Integrating over $\boldsymbol{\rho}_1$ gives

$$\iint \psi_{dG}^*(\boldsymbol{\rho}_1, \boldsymbol{\rho}_2) \psi_{dG}(-\boldsymbol{\rho}_1, \boldsymbol{\rho}_2) d^2 \boldsymbol{\rho}_2 \, d^2 \boldsymbol{\rho}_1, \quad (A24)$$

the inverse of which is equal to the Schmidt number. Changing variables from $(\boldsymbol{\rho}_1, \boldsymbol{\rho}_2)$ to $(\boldsymbol{\rho}_+, \boldsymbol{\rho}_-)$ transforms the wave function according to $\psi(\boldsymbol{\rho}_1, \boldsymbol{\rho}_2) \to \psi(\boldsymbol{\rho}_+, \boldsymbol{\rho}_-)$ and $\psi(-\boldsymbol{\rho}_1, \boldsymbol{\rho}_2) \to \psi(-\boldsymbol{\rho}_-, -\boldsymbol{\rho}_+)$

yielding Eq. (6):

$$K_{dG} = \left[\iint \psi_{dG}^*(\boldsymbol{\rho}_+, \boldsymbol{\rho}_-)\psi_{dG}(-\boldsymbol{\rho}_-, -\boldsymbol{\rho}_+)d^2\boldsymbol{\rho}_+ d^2\boldsymbol{\rho}_-\right]^{-1}. \quad (A25)$$

*mr22@princeton.edu
†jasonf@princeton.edu